\documentclass[prb,twocolumn,showpacs,superscriptaddress,preprintnumbers]{revtex4}

\usepackage{graphicx}
\usepackage{dcolumn}
\usepackage{bm}

\begin{document}

\preprint{APS/123-QED}

\title{Structure-Property Relationship in the Ordered-Perovskite- Related Oxide Sr$_{3.12}$Er$_{0.88}$Co$_{4}$O$_{10.5}$
}

\author{Shintaro Ishiwata}
 \email{ishiwata@htsc.sci.waseda.ac.jp}
 \affiliation{Department of Applied Physics, Waseda University, Shinjuku, Tokyo 169-8555, Japan.}

\author{Wataru Kobayashi}%
\affiliation{Department of Applied Physics, Waseda University, Shinjuku, Tokyo 169-8555, Japan.}%

\author{Ichiro Terasaki}
\affiliation{Department of Applied Physics, Waseda University, Shinjuku, Tokyo 169-8555, Japan.}%

\author{Kenichi Kato}
\affiliation{RIKEN/SPring-8 Center, Sayo, Hyogo 679-5148, Japan.}
\affiliation{Japan Synchrotron Radiation Research Institute (JASRI), Sayo, Hyogo 679-5198, Japan.}%
\affiliation{CREST,Japan Science and Technology Agency, Kawaguchi, Saitama 332-0012, Japan.}%

\author{Masaki Takata}
\affiliation{RIKEN/SPring-8 Center, Sayo, Hyogo 679-5148, Japan.}
\affiliation{Japan Synchrotron Radiation Research Institute (JASRI), Sayo, Hyogo 679-5198, Japan.}%
\affiliation{CREST,Japan Science and Technology Agency, Kawaguchi, Saitama 332-0012, Japan.}%

\date{\today}

\begin{abstract}
Synchrotron X-ray diffraction patterns were measured and analyzed for a polycrystalline sample of the room-temperature ferromagnet Sr$_{3.12}$Er$_{0.88}$Co$_{4}$O$_{10.5}$ from 300 to 650 K, from which two structural phase transitions were found to occur successively. The higher-temperature transition at 509 K is driven by ordering of the oxygen vacancies, which is closely related to the metallic state at high temperatures. The lower-temperature transition at 360 K is of first order, at which the ferromagnetic state suddenly appears with exhibiting a jump in magnetization and resistivity. Based on the refined structure, possible spin and orbital models for the magnetic order are proposed.
 
\end{abstract}

\maketitle
Among various transition-metal oxides, trivalent cobalt oxides exhibit unique electronic phases characterized by the interplay between nearly degenerate spin states. For example, the Co$^{3+}$ ions in LaCoO$_{3}$ are in the nonmagnetic low-spin (LS) state ($t_{2g}^{6}$, $S$ = 0) at low temperatures, and are thermally excited to be magnetic with increasing temperature\cite{Heikes, Naiman, Raccah}. Although the most stable magnetic state is the high-spin (HS) state ($t_{2g}^{4}e_{g}^{2}$, $S$ = 2), the intermediate-spin (IS) state ($t_{2g}^{5}e_{g}^{1}$, $S$ = 1) is believed to exist in intermediate temperature range \cite{Korotin, Yamaguchi, Asai}. Since octahedrally coordinated Co$^{3+}$ ions in the IS state have orbital degree of freedom, an orbital order may occur to lift the degeneracy of the partially filled $e_{g}$ orbitals \cite{Korotin, Zobel, Ishikawa}.  Actually, Maris et al. found a trace of the orbital order by X-ray diffraction studies \cite{Maris}.  Having said that, we think that the IS state in LaCoO$_{3}$ is still controversial \cite{Pod}, because it appears as a transient state at moderately high temperatures.

Recently, an unusual ferromagnetism has been found in the new trivalent Co oxide Sr$_{3}$RCo$_{4}$O$_{10.5}$ (R = Y, Ho, Er, etc.) \cite{Kobaprb} having a spontaneous magnetization below 330-360 K with an insulating ground state. In spite of the remarkably high Curie temperature, the ferromagnetic order is fragile against a small substitution of Mn for Co, in which only 6 \% Mn quenches the spontaneous magnetization \cite{Koba}. This strongly suggests that the origin of the ferromagnetism is highly unconventional. Since the ferromagnetic state is the ground state of this oxide, it offers a unique opportunity to study the IS/HS state as a long-range-ordered state, which can be quite precisely investigated with diffraction techniques.

Here we show a peculiar structure-property relationship by measuring X-ray diffraction and physical properties for a polycrystalline sample of Sr$_{3.12}$Er$_{0.88}$Co$_{4}$O$_{10.5}$ from 300 to 650 K. Two structural transitions are found to occur successively in this temperature range: The first transition occurs at 509 K, which is closely related to the metallic state at high temperatures. In contrast, the second transition observed at 360 K is coupled with the ferromagnetic order. From detailed structure analyses, we propose possible origins for the ferromagnetism.

Polycrystalline samples Sr$_{3.12}$Er$_{0.88}$Co$_{4}$O$_{12- \delta}$ were obtained by solid-state reaction in air as described in ref. [11]. The nominal composition was Sr$_{0.78}$Er$_{0.22}$Co$_{1.05}$O$_{3-y}$, in which a 5 \% excess of Co$_{3}$O$_{4}$ was added to compensate evaporation during the reactions. To minimize unwanted intermixture between the Sr and Er ions, the Er/Sr ratio was slightly shifted to the Sr-rich side. X-ray diffraction (XRD) data were collected with a wavelength of 0.77667 $\rm{\AA}$ on an imaging plate using a large Debye-Scherrer camera installed at BL02B2, SPring-8 \cite{Takata}. A well-ground polycrystalline sample was sealed into a glass capillary with an internal diameter of 0.2 mm. Temperatures were stabilized within 1 K by a N$_{2}$ gas flow apparatus. Structural analyses were performed by a Rietveld refinement using a RIETAN-2000 program \cite{Izumi}, using the data in 3.5 deg $<$ 2$\theta$ $<$ 70 deg with 0.01 deg step. The equivalent isotropic thermal factors for the same species were constrained to be equivalent during the refinements. The Er/Sr ratio (= 0.88/3.12) in the A site and the amount of oxygen vacancy $\delta$ (= 1.5) were refined using the 650-K data. Since a cubic perovskite phase \cite{Istomin} coexists as a secondary phase (the volume fraction of approximately 10 \%), the structure refinements were performed by two-phase profile analyses. DC magnetic susceptibility was measured with a MPMS SQUID magnetometer (Quantum Design) in an external magnetic field of 0.1 T.  Electrical resistivity was measured by a four-probe method in an electronic furnace. 

\begin{figure}
\includegraphics[keepaspectratio,width=6.5cm]{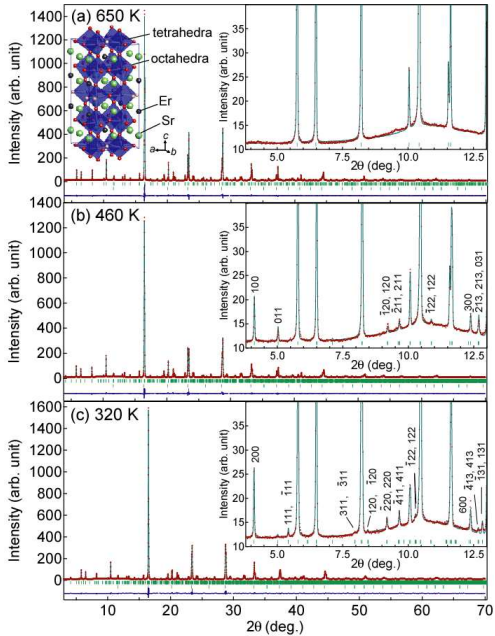}
\caption{\label{fig1} (Color online) Observed (+), calculated (solid line), and differential Rietveld refinement X-ray diffraction profiles of Sr$_{3.12}$Er$_{0.88}$Co$_{4}$O$_{10.5}$ at (a) 650 K, (b) 460 K, and (c) 320 K. The low angular region (3.8 deg $\leq$ 2$\theta$ $\leq $ 13.0 deg) was enlarged and shown as the insets.  The tick marks indicate the calculated peak positions.  The crystal structure at 650 K is shown in the inset.
}
\end{figure}

Figure 1 shows representative XRD patterns of Sr$_{3.12}$Er$_{0.88}$Co$_{4}$O$_{10.5}$ with calculated profiles at selected temperatures.  The diffraction pattern at 650 K was fitted with a body-centered tetragonal unit cell ($I$4/$mmm$) of 2$a_{\rm 0}$ $\times$ 2$a_{\rm 0}$ $\times$ 4$a_{\rm 0}$, where $a_{\rm 0}$ is the cell length of the primitive perovskite unit cell.  As illustrated in Fig. 1(a), the A-site ions are ordered, and the Co-O network forms a brownmillerite-like layered structure consisting of an alternate stack of the CoO$_{6}$ octahedral layers and the CoO$_{4.25}$ tetrahedral layers. This structure is essentially the same as was previously reported independently by Istomin et al. \cite{Istomin2} and by Withers et al. \cite{Withers}

We have discovered two structural phase transitions in the measured temperature range. The insets of Fig. 1(b) and 1(c) have clearly revealed superstructure peaks, and indicate that the Co-O network is further ordered with larger periodicities of 2$\sqrt{2}a_{\rm 0} \times 2\sqrt{2} a_{\rm 0} \times 4a_{\rm 0}$ at 460 K and 4$\sqrt{2}a_{\rm 0} \times 2\sqrt{2} a_{\rm 0} \times 4a_{\rm 0}$ at 320 K.  These superstructure peaks are weaker than the main peak by three orders of magnitude, which could not have been detected without a high-intensity synchrotron X-ray source.  In addition, some Bragg reflections thus far indexed as the tetragonal symmetry are split, which indicates that the ordered structure is monoclinic.  The extinction rules obtained from the diffraction patterns are $k$+$l$ = 2$n$ for $hkl$ and 0$kl$, $k$ = 2$n$ for $hk$0 and 0$k$0, and $l $= 2$n$ for $h$0$l$ and 00$l$, indicating that a possible space group is $A$2/$m$, $Am$ or $A$2.  Since attempts to fit the data with $Am$ and $A$2 did not improve the refinement, we have adopted $A$2/$m$ ---the most symmetric one among them. The refined lattice parameters and the reliability factors thus obtained are given in Table I. In spite of the complicated unit cell, the calculated profiles and the reliability factors are satisfactory. We call the phase at 650 K the high-temperature tetragonal (HTT) phase. Similarly, we name the phases at 460 K and 320 K the high-temperature monoclinic (HTM) phase and the low-temperature monoclinic (LTM) phase, respectively.

\begin{figure}
\includegraphics[keepaspectratio,width=7.4cm]{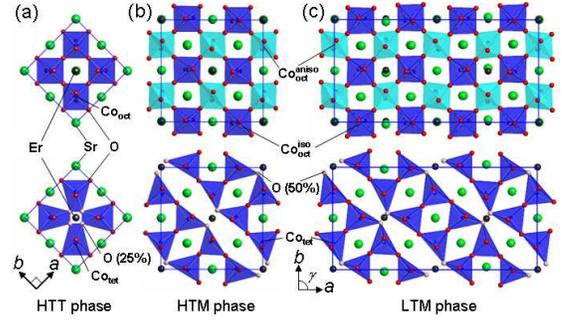}
\caption{\label{fig2} (Color online) Crystal structures of the CoO$_{6}$ layer (upper) and the CoO$_{4.25}$ layer (lower) at (a) 650 K, (b) 460 K, and (c) 320 K, viewed along the $c$ axis.  The light red (light gray) spheres in the the CoO$_{4.25}$ layers denote oxygen sites with an occupancy of 25 $\%$ for (a) and an occupancy of 50 $\%$ for (b) and (c). 
}
\end{figure}

In Fig. 2, the CoO$_{6}$ layer and the CoO$_{4.25}$ layer for the three phases are schematically illustrated by viewing along the $c$ axis. Oxygen-vacancy ordering in the CoO$_{4.25}$ layer is found to drive the HTT-HTM transition from a careful comparison of Figs. 2(a) and 2(b). The one extra oxygen (denoted by 0.25 in CoO$_{4.25}$) randomly occupies one out of the four positions in Fig. 2(a), whereas it occupies one out of the two in Fig. 2(b). The CoO$_{6}$ and the CoO$_{4.25}$ polyhedra are slightly tilted and distorted to accept the atomic displacements due to the vacancy ordering in the CoO$_{4.25}$ layer. In contrast, comparing Fig. 2(c) with Fig. 2(b), we see that the significant distortion of the CoO$_{6}$ octahedra and the concomitant displacements of A-site ions predominantly drive the HTM-LTM transition, accompanied with cell doubling along the $a$ axis. 

\begin{table*}
\caption{\label{table1}Lattice parameters, space group, and reliability factors for HTT (650 K), HTM (460 K), and LTM (320 K) phases.}
\begin{ruledtabular}
\begin{tabular}{ccccccccc}
$T$ (K) & S.G. & $a$ (\AA) & $b$ (\AA) & $c$ (\AA) & $\gamma$ (deg) &$V (\rm \AA^{3})$ & $R_{\rm WP}$(\%) &$R_{\rm I}$(\%)  \\\hline
650 & $I$4/$mmm$ & 7.67256(3) & 7.67256(3) & 15.4527(1) &  & 909.672(8) & 4.49 & 1.91 \\
460 & $A$2/$m$ & 10.80867(8)   & 10.82760(9) & 15.4007(1)  & 90.1510(6) & 1802.37(2) & 4.67 & 1.8 \\
320 & $A$2/$m$ & 21.5942(2)   & 10.8461(2) & 15.3481(2) & 90.075(2) & 3594.69(8) & 5.09 & 1.25 \\
\end{tabular}
\end{ruledtabular}
\end{table*}

Figures 3(a) and 3(b) show the lattice parameters plotted as a function of temperature, from which the two transition temperatures are determined to be $T_{\rm c1}$=509 K (the HTT-HTM transition) and $T_{\rm c2}$=360 K (the HTM-LTM transition). Regarding the HTT-HTM transition, as we detect no discontinuities in the structural variables, the transition appears to be approximately continuous. The monoclinic angle $\gamma$ for the HTM phase was excellently expressed by the equation of $\gamma$  = 90 + 0.011($T_{\rm c1}$-$T$)0.67 [deg], which suggests the second-order nature of the transition at $T_{\rm c1}$. The diffuse reflection of 1/2 1/2 0 [corresponding to the 100 reflection for the HTM phase] is observed in the XRD patterns above $T_{\rm c1}$ up to 600 K (the inset of Fig. 3), which is further suggestive of the second-order structure transition. In contrast, the transition at $T_{\rm c2}$ is of first order, at which the lattice parameters discontinuously change. 

In order to see the structure-property relationship in Sr$_{3.12}$Er$_{0.88}$Co$_{4}$O$_{10.5}$, we measured magnetic susceptibility and resistivity of the same sample as shown in Figs. 3(c) and 3(d). Reflecting the first-order transition, the inverse susceptibility $H/M$ and the resistivity $\rho$  change discontinuously across $T_{\rm c2}$. On the other hand, the change of both quantities across $T_{\rm c1}$ is continuous. We should note here that the resistivity becomes metallic (i.e. d$\rho$/d$T > $0) above 620 K \cite{Kobajpsj}, which roughly coincides with the disappearance of the 1/2 1/2 0 diffuse scattering. This implies that formation of short-range-order domains of the HTM phase is quite effective to localize carriers on the magnetic Co$^{3+}$  ions. Considering that the oxygen-vacancy ordering leads to the insulating state, the CoO$_{4.25}$ layer is unlikely to be in charge of a good conductivity, because the lattice of CoO$_{4.25}$ layer is more disordered above $T_{\rm c1}$. Consequently, the CoO$_{6}$ layer should be responsible for the metallic state above 620 K.

\begin{figure}
\includegraphics[keepaspectratio,width=5.5cm]{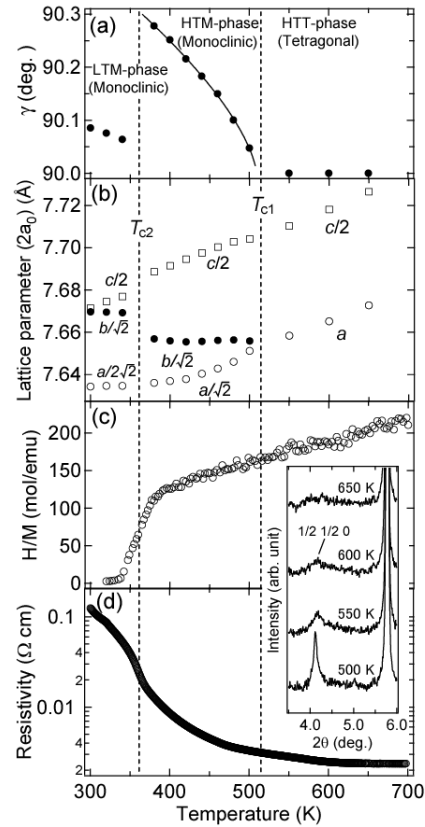}
\caption{\label{fig3} Temperature dependence of (a) monoclinic angle, (b) axis lengths, (c) inverse molar magnetic susceptibility, and (d) resistivity of Sr$_{3.12}$Er$_{0.88}$Co$_{4}$O$_{10.5}$.  The axis lengths are given for 2$a_{\rm 0}$ ($a_{\rm 0}$ is the lattice parameter for the simple cubic perovskite cell).  The inset shows development of the 1/2 1/2 0 superstructure reflection in the HTT phase.  
}
\end{figure}

Reflecting the complicated orders, the inequivalent Co sites drastically increase with decreasing temperature. For the HTT, HTM and LTM phases, the numbers of the inequivalent Co sites in the CoO$_{6}$ layer are 1, 2 and 5, while those in the CoO$_{4.25}$ layer are 1, 4 and 8, respectively. For the sake of simplicity, we will neglect the small difference in the inequivalent sites of the CoO$_{4.25}$ layer, and symbolize them by Co$_{\rm tet}$  as indicated in Fig. 2(a). Similarly, we will classify the inequivalent sites in the CoO$_{6}$ layer into Co$_{\rm oct}^{\rm iso}$  and Co$_{\rm oct}^{\rm aniso}$ , as indicated in Figs. 2(b) and 2(c). Figure 4(a) shows temperature dependence of the averaged Co-O bond lengths for Co$_{\rm tet}$  and Co$_{\rm oct}$ sites, in which those for Co$_{\rm oct}^{\rm iso}$  and Co$_{\rm oct}^{\rm aniso}$  sites are further averaged.  With decreasing temperature from 650 to 420 K, the averaged bond length for the Co$_{\rm oct}$ sites decreases, whereas that for Co$_{\rm tet}$  sites monotonically increases owing to the oxygen vacancy ordering. Note that the Co-O length for the Co$_{\rm oct}$ sites is significantly (2-3 \%) larger than that of LaCoO$_{3}$ \cite{Thornton}. Since the Co$^{3+}$  ions in LaCoO$_{3}$ are believed to be in the IS or HS state in this temperature range, the Co$^{3+}$ ions in the Co$_{\rm oct}$  sites are also expected to be in the IS/HS state above 300 K.  

Next, we show the differences in the Co-O bonds in the Co$_{\rm oct}$  sites. As seen in Fig. 4(b), the Co-O bond lengths for the Co$_{\rm oct}^{\rm aniso}$  sites split into three different values below $T_{\rm c1}$, whereas the Co-O bond lengths for the Co$_{\rm oct}^{\rm iso}$  sites tend to come closer to each other. At the HTM-LTM transition, the Co$_{\rm oct}^{\rm aniso}$ -O bond lengths are drastically changed, whereas the change for the Co$_{\rm oct}^{\rm iso}$ -O bonds is subtle. 

Let us propose two possible models for the magnetic order in Sr$_{3.12}$Er$_{0.88}$Co$_{4}$O$_{10.5}$ on the basis of the refined structure (see Fig. 4(c)). The first one is that the Co$_{\rm oct}^{\rm aniso}$  and Co$_{\rm oct}^{\rm iso}$  ions are in the IS and HS states below $T_{\rm c2}$, respectively. In this structure, all of the nearest-neighbor interactions are antiferromagnetic, which gives a ferrimagnetic state composed of the antiferromagnetically coupled HS ($S$ = 2) and IS ($S$ = 1) states.  The second model is that all the Co ions are in the IS state. Then, an orbital order is realized in such a way that the $x^{2}$-$z^{2}$/$y^{2}$-$z^{2}$ orbitals pointing toward $<$110$>$ or $<\bar{1}$10$>$ reside at the Co$_{\rm oct}^{\rm aniso}$  sites and the $z^{2}$-$r^{2}$ orbitals pointing toward $<$001$>$ reside at the Co$_{\rm oct}^{\rm iso}$  sites. Accordingly, ferromagnetic zigzag chains composed of partially occupied $x^{2}$-$z^{2}$/$y^{2}$-$z^{2}$ orbitals at the Co$_{\rm oct}^{\rm aniso}$  sites hybridizing with unoccupied $x^{2}$-$y^{2}$ orbitals at the Co$_{\rm oct}^{\rm iso}$  sites are formed along the $a$ axis. This orbital order is similar to the CE-type orbital order proposed for La$_{0.5}$Ca$_{0.5}$MnO$_{3}$ .\cite{Goodenough, Chen} Since an interaction between the ferromagnetic chains is expected to be antiferromagnetic, the ferromagnetic moment in the second model stems from a spin canting due to the distorted structure. In both models, an orbital ordering of Co$_{\rm oct}^{\rm aniso}$  in the IS state plays a crucial role for occurrence of the cooperative lattice distortion leading to the cell doubling along the $a$ axis.  Although CoO$_{4.25}$ layers may favor the IS or HS state owing to the small crystal field splitting, they are unlikely to be magnetic. Preliminarily, we confirmed that the ferromagnetic order is robust against partial substitution of Al for Co in the tetrahedral site. Given the nonmagnetic CoO$_{4.25}$ layers, the ferrimagnetic spin structure in the first model gives the moment of 0.5 $\mu_{\rm B}$/Co, while the canted spin structure in the second model affords the moment of 1 $\mu_{\rm B}$/Co at maximum.

\begin{figure}[t]
\includegraphics[keepaspectratio,width=7.2cm]{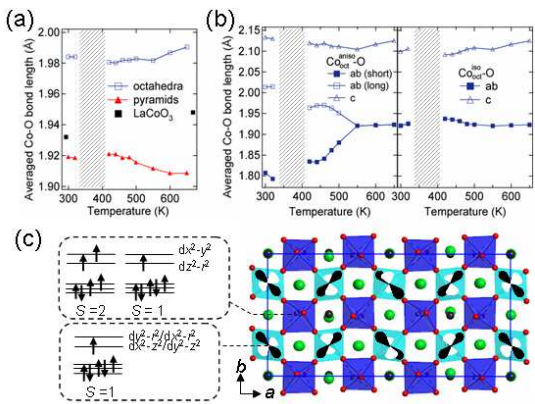}
\caption{\label{fig4} (Color online) (a) Averaged Co-O bond lengths for the octahedral and the tetrahedral sites of Sr$_{3.12}$Er$_{0.88}$Co$_{4}$O$_{10.5}$. Averaged Co-O bond lengths of LaCoO$_{3}$ are taken from ref. [19]. (b) Averaged Co-O bond lengths for Co$_{\rm oct}^{\rm aniso}$  (left panel) and that for Co$_{\rm oct}^{\rm iso}$  (right panel). Typical errors are $\sim$0.02 \AA.  The Co-O bond lengths in the shaded area are not displayed because of poor reliability of the refinements.  (c) Possible spin and orbital configurations at Co$_{\rm oct}^{\rm aniso}$ and Co$_{\rm oct}^{\rm iso}$  in the LTM phase.
}   
\end{figure}

The complicated structure in the LTM phase explains various anomalous properties in this compound. First, since the HTM-LTM transition is driven by a ferrimagnetic or a canted antiferromagnetic ordering, the transition temperature can be as high as the Neel temperature of other Co oxides, energy scale of which is the Co-O-Co super-exchange interaction. Second, the ordered states proposed here explain the small saturation moment of 0.5 $\mu_{\rm B}$/Co. Third, the large unit cell of the LTM phase explains why the ferromagnetic order is sensitive against Mn doping. The spin/orbital state of CoO$_{6}$ layers in the LTM phase is barely stable with the coherent displacement of the 64 Co ions. Thus, when one of the 32 Co ions in the CoO$_{6}$ layer of the unit cell is replaced by a Mn ion, the rest of 31 Co ions will lose magnetism. If Mn ions are substituted equally in the CoO$_{6}$ and CoO$_{4.25}$ layers, the critical concentration will be around 1/32, which is close to the actually measured value of 6 $\%$. 

In summary, we have found two successive structural transitions in Sr$_{3.12}$Er$_{0.88}$Co$_{4}$O$_{10.5}$ between 300 and 650 K.  At $T_{\rm c1}$= 509 K, the oxygen-vacancy ordering gives rise to the tetragonal-monoclinic phase transition of second order, which induces the cooperative rotation of Co-O polyhedra. The first order transition at $T_{\rm c2}$= 360 K appears to be driven by a spin/orbital ordering in the octahedral layer, where Co ions in the IS state play an important role for the room-temperature ferromagnetism. Transmission-electron-microscope studies are necessary to examine the space group and local structures, which should be done as future works.

The authors thank A. A. Belik, M. Takano, H. Nakao, and Y. Murakami for useful discussions, and T. Iizuka for collaboration on the Co-site doping effects. This work was supported by the Academic Frontier Project from the MEXT and the 21$^{\rm st}$ Century COE Programs at Waseda University for Physics, and the Grant-in-Aid for Scientific Research (No. 16076213). The synchrotron radiation experiments were performed under the Program No. 2006A1798.  S. I. receives additional supports from JSPS.

\appendix

\bibliography{apssamp}

\end{document}